\documentclass[aps,prl%
,twocolumn%
,showpacs%
]{revtex4}

\usepackage{graphicx}
\usepackage{times}

\begin{document}

\setlength{\textfloatsep}{1.2\parindent plus 2pt minus 2pt} %%% DELETE %%%
\setlength{\arraycolsep}{1pt}
\newcommand{\be}{\begin{equation}}
\newcommand{\ee}{\end{equation}}
\newcommand{\bea}{\begin{eqnarray}}
\newcommand{\eea}{\end{eqnarray}}
\newcommand{\f}{\frac}
\newcommand{\p}{\partial}
\newcommand{\no}{\nonumber}
\newcommand{\kT}{k_{\rm B}T}
\newcommand{\e}{{\rm e}}
\newcommand{\dd}{{\rm d}}

\newcommand{\kav}{\left<k\right>}

\newcommand{\affila}{
 Biological Physics Research Group of HAS and
 Department of Biological Physics, E\"otv\"os University,
 P\'azm\'any P.\ stny.\ 1A, H-1117 Budapest, Hungary
}
\newcommand{\affilb}{
}

\title{
Topological phase transitions of random networks
}

\author{Imre Der\'enyi}
 %\email{derenyi@elte.hu}
 \affiliation{\affila}
\author{Ill\'es Farkas}
 %\email{fij@elte.hu}
 \affiliation{\affila}
\author{Gergely Palla}
 %\email{pallag@angel.elte.hu}
 \affiliation{\affila}
\author{Tam\'as Vicsek}
 %\email{vicsek@angel.elte.hu}
 \affiliation{\affila}

\date[]{\protect\today}
%\date[]{submitted to where..., when...}

\begin{abstract}
To provide a phenomenological theory for the various interesting transitions in restructuring networks we employ a statistical mechanical
approach with detailed balance satisfied for the transitions
between topological states. This enables us to establish an equivalence
between the equilibrium rewiring problem we consider and the dynamics of a lattice gas on
the edge-dual graph of a fully connected network. By assigning
energies to the different network topologies and defining the
appropriate order parameters, we find a rich variety of topological
phase transitions, defined as singular changes in the essential feature(s) of the global connectivity as a
function of a parameter playing the role of the temperature. In the
``critical point'' scale-free networks can be recovered.
\end{abstract}


%89.75.Hc Networks and genealogical trees
%05.70.Fh Phase transitions: general studies
%64.60.Cn Order-disorder transformations; statistical mechanics of model systems
%87.23.Ge Dynamics of social systems

\maketitle

One of the most fruitful recent approaches to the description of
complex systems consisting of many similar units has been the analysis
of the underlying network of interactions. It has turned out that
seemingly very different systems can be characterized by a few major
classes of stochastic graphs representing the overall features of the
structure of connections among the units
\cite{b-a-revmod,dorog-mendes-book}. These developments have greatly
advanced our potential to interpret the fundamental common features of
diverse systems including social groups, technological, biological and
other networks. The effects of both the restructuring
\cite{watts2} and
the growth \cite{b-a-science} of the associated graphs have been
considered leading to a number of exciting discoveries about the laws
concerning their diameter, clustering and degree distribution.

Although the conceptual basis of most of the related works has largely
relied on analogies (scaling, percolation, etc.) with statistical
mechanics, only very few recent works
\cite{burda,berg,burda2} have been
devoted to the problem of directly connecting the graph theoretical
aspects of networks to statistical mechanics or thermodynamics. On the
other hand, using a thermodynamic formalism for the changes
in graphs being in an equilibrium-like state is expected to provide a
significantly deeper insight into the processes taking place in systems
being in a saturated state and, as such, dominated by the fluctuating
rearrangements of links between their units.

As an example, let us take a given number of units interacting in a
``noisy'' environment. These units can be people, firms, genes, etc. The
probability for establishing a new or ceasing an existing
interaction/connection between two units depends on both the noise and
the advantage/disadvantage gained/lost when adopting the new
configuration. 
In this picture, a global transition in the connectivity properties can occur as a function
of the level of perturbations. For instance, if the conditions are such
that the interactions between the partners become more ``conservative''
(a relative, short term gain is more highly valued), as we show
later, a transition from a less ordered to a more ordered network configuration can take place. In particular, it has been argued \cite{stark-vedres} that depending on the level of given types of   uncertainties (expected fluctuations) business networks reorganize from a star-like topology to a system of more cohesive, highly clustered ties. 

In this paper we use temperature to represent noise and differences in an energy (potential) type quantity to  
account for advantage or loss during the
rearrangement of a network. This enables us to treat the associated
microscopic dynamics within the framework of {\it canonical ensembles}
and map the problem onto a {\it lattice gas} model.
After defining an order parameter and a variety of possible potentials,
interesting {\it topological phase transitions (first and second order)} are found both analytically
and by numerical simulations. Scale-free networks are recovered at the
transition point between a system consisting of many nodes with few links
to a system dominated by a single large hub.

We define the partition function
\be
Z(T)=\sum_{ \{ g_a \} } \e^{-E_a / T}
\label{Z}
\ee
for an ensemble $\{ g_a \}$ of undirected graphs
containing $N$ nodes and $M$ links,
where $E_a$ denotes the energy of the graph $g_a$.
Our choice for $Z$ is in
analogy with that proposed by Berg and L\"assig
\cite{berg} and is motivated by the
following physical picture behind the graph restructuring process: The
basic event of the rearrangement is the relocation of a randomly
selected edge (link) to a new position either by ``diffusion'' (keeping
one end of the edge fixed and connecting the other one with a new node)
or by removing the given edge and connecting two randomly selected
nodes. Then, the energy difference
$\Delta E_{ab}=E_b-E_a$
between the original $g_a$ and the new $g_b$ configurations is
calculated and the relocation is carried out
following the Metropolis algorithm.
The resulting dynamics, by construction, satisfies the detailed balance
condition.

This network rearrangement is formally equivalent to a {\em
Kawasaki type lattice gas} dynamics with conserved number of particles
moving on a special lattice, which is the edge-dual graph of the fully
connected network
\cite{bollobas,ramezanpour}.
The sites of this lattice are the possible
$N(N-1)/2$ connections between the vertices, and the particles wandering
on the sites are the $M$ edges. 

To be able to monitor topological phase transitions a suitable {\it order
parameter} $\Phi$ has to be introduced. As we are primarily
interested in the transitions between dispersed and compact states, a
natural choice can be either
$\Phi=\Phi_s=s_{\rm max}/M$, the number of edges of the largest
connected component of the graph
$s_{\rm max}$ normalized by the total number of edges $M$, or
$\Phi=\Phi_k=k_{\rm max}/M$, the highest degree in the graph
$k_{\rm max}$ divided by $M$. We also introduce the
corresponding conditional free energy $F(\Phi,T)$ via
\be
\e^{-F(\Phi,T) / T}=Z(\Phi,T)=\sum_{ \{ g_a \}_\Phi } \e^{-E_a / T},
\label{FPhi}
\ee
where $\{ g_a \}_\Phi$ is a subset of $\{ g_a \}$ and contains the
graphs with the same order parameter $\Phi$.

In the $T\to\infty$ limit 
the dynamics converges to a totally random rewiring
process, and thus, the classical Erd\H{o}s-R\'enyi (ER)
\cite{ErdosRenyi}
random graphs are recovered. On the other hand, at low temperatures the
topologies with lowest energy occur with enhanced probability. If the
energy function is chosen such that compact configurations (which occur
exceptionally rarely in the ER graphs) have low energies, then in an
intermediate temperature range we expect to see a competition between
the entropically favorable dispersed configurations and the
energetically favorable compact ones. This is manifested in the shift
of the minimum of the conditional free energy $F(\Phi,T)$ from $\Phi=0$
towards higher values of $\Phi$ as the temperature $T$ is decreased
from infinity to zero. A sudden change in the position of the global
minimum signals a discontinuous (first order) phase transition, whereas a
gradual shift indicates either a cross-over or a continuous (second
order) phase transition.

In the classical random graph model \cite{bollobas} (corresponding to $T\to\infty$)
by varying the average degree of the vertices $\kav=2M/N$, a
percolation phase transition occurs at $\kav=1$. For
$\kav<1$ the graphs falls apart into small pieces, while
for $\kav\geq 1$ a giant connected component emerges. Near
the critical point the size of this giant component scales as
$(\kav-1)M$.

Based on the lattice gas analogy we expect that if $\kav<1$, then for a
suitable choice of the energy (one that rewards clustering) a similar
dispersed-compact phase transition occurs at a finite temperature
$T(\kav)$. Such a transition can be best monitored by the order
parameter
$\Phi_s=s_{\rm max}/M$ \cite{dorog-mendes-book,ErdosRenyi}.

The most obvious energy satisfying the above requirement is a
monotonically decreasing function $E=f(s_{\rm max})$.
It can be shown (by counting the number of configurations) that the
conditional free energy up to second order in $\Phi_s$ 
(and after omitting the $\Phi_s$ independent terms) can
be approximated by
\bea
\f{F(\Phi_s,T)}{MT} &\approx&
 \f{f(\Phi_s M)}{MT}
 +\left[\kav-1-\ln(\kav)\right]\Phi_s
\no\\
 &&+\left[\kav^2-3\kav+2\right]\f{\Phi_s^2}{4}
\label{Fs}
\eea
in the $N,M\to\infty$ limit for fixed $\kav=2M/N$.

\begin{figure}[t!]
\centerline{\includegraphics[angle=-90,width=0.9\columnwidth]{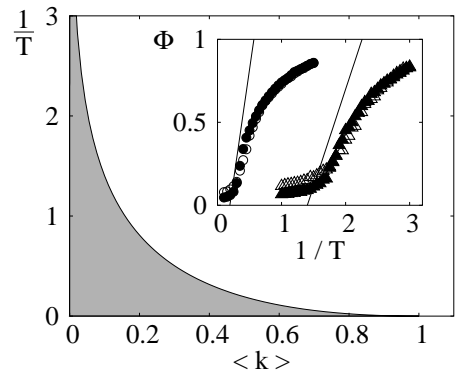}}
\caption{
The phase diagram and the order parameter for the $E=-s_{max}$ energy.
Main panel: The white and shaded areas correspond to the ordered phase
(containing a giant component) and the disordered phase, respectively.
Inset: The order parameter $\Phi=\Phi_s=s_{\rm max}/M$
obtained from Monte-Carlo simulations
as a function of the inverse
temperature for $\langle k\rangle=0.1$ (triangles) and 
$\langle k\rangle=0.5$ (circles).
Each data point is an ensemble average of $10$ runs, time averaged
between $t=100N$ and $500N$ Monte-Carlo steps. The open and closed
symbols represent $N=500$ and $1,000$ vertices, respectively.
The critical exponent, in agreement with the
analytical approximations (solid lines),
was found to be $1$.
}
\label{fig_s}
\end{figure}

The simplest choice for the energy function is
$f(s_{\rm max})=-s_{\rm max}$. In this case it can be clearly seen from
Eq.\ (\ref{Fs}) that as long as
$1/T<1/T_{\rm c}(\kav)=\kav-1-\ln(\kav)$,
the free energy has a minimum at
$\Phi_s=\Phi_s^*(T)=0$, i.e., the configuration is
dispersed (see main panel of Fig.\ \ref{fig_s}). 
When the temperature drops below
$T_{\rm c}(\kav)$, the minimum moves away from $\Phi_s=0$ and a giant
component appears. Near the critical temperature $T_{\rm c}(\kav)$ the
order parameter at the minimum of the free energy can be estimated from
Eq.\ (\ref{Fs}) as
$\Phi_s^*(T)=2[1/T-1/T_{\rm c}(\kav)]/[\kav^2-3\kav+2]$,
indicating that we are dealing with a {\it second order topological phase
transition} (see inset of Fig.\ \ref{fig_s}).

For other forms of $f(s_{\rm max})$, such as $-s_{\rm max}^2$ or
$-s_{\rm max}\ln(s_{\rm max})$, first order topological phase
transitions are also expected to occur. We have found such transitions
numerically (not shown).
In addition, similar results are expected for any (non-global) energy
$E=\sum_j f(s_j)$, where the summation goes over each connected
component and $s_j$ denotes the number of edges in the $j$th one.
However, because the total number of edges, $M=\sum_j s_j$, is
conserved by the dynamics, $f(s_j)$ must decrease faster than $-s_j$
in this case.

Next we turn to another important class of the energy functions, where the
energies are assigned to the vertices rather than to the connected
components of the graph:
$E=\sum_{i=1}^{N}f(k_i)$,
where $k_i$ denotes the degree (number of neighbors) of vertex $i$.
This energy is consistent with a dynamics, in which the change of the
degree of a vertex depends only on the structure of the graph in its
vicinity. The fitness
of an individual vertex depends on its
connectivity. The most suitable order parameter for this class of
graph energy is
$\Phi_k=k_{\rm max}/M$.
Again, due to the conservation of the number of edges,
$M=\sum_i k_i$, the single vertex energy $f(k_i)$ should decrease
faster than $-k_i$, if aggregation is to be favored.

If we express the single vertex energy in the form
$f(k_i)=k_i g(k_i)$,
then the total energy of the graph can also be written as
$E=\sum_{i=1}^{N}f(k_i)=\sum_{i=1}^{N}\sum_{i'}g(k_{i'})$,
where $i'$ runs over all vertices that are neighbors of vertex $i$.
Therefore, the graph energy can also be interpreted in such a way, that
every vertex $i$ collects an energy $g(k_{i'})$ from each of its
neighbors. In this interpretation, the fitness of an individual vertex depends on
the connectivities of its neighbors.

A relevant case is when $f(k_i)= -(J/2)k_i^2$, or equivalently,
$g(k_i)=-(J/2)k_i$, corresponding to a linear preference of the number
of pairs that can be chosen from the edges of a vertex, or the number
of edges owned by the neighbors of a vertex.
Furthermore, this form is in full analogy with the usual
definition of the  
energy
$E=-J\sum_{<i,j>} n_i n_j$
of a lattice gas on the edge-dual graph of the fully connected network
with nearest neighbor attraction. The summation here runs over all
adjacent pairs of lattice sites, and $n_i=1$ if site $i$ is occupied
and $0$ otherwise. When this energy is applied to normal cubic
lattices, we recover the standard lattice gas model of nucleation of
vapors. The negative energy unit $-J$ associated with a pair of edges
sharing a vertex in the original graph is equivalent to the binding
energy between the corresponding occupied nearest neighbor sites on the
edge-dual graph. By measuring the energies (and temperature) in units
of $J$ we can set $J=1$, without loosing generality. Thus, from now on
$J$ will be omitted.

\begin{figure}[t!]
\centerline{\includegraphics[angle=-90,width=0.9\columnwidth]{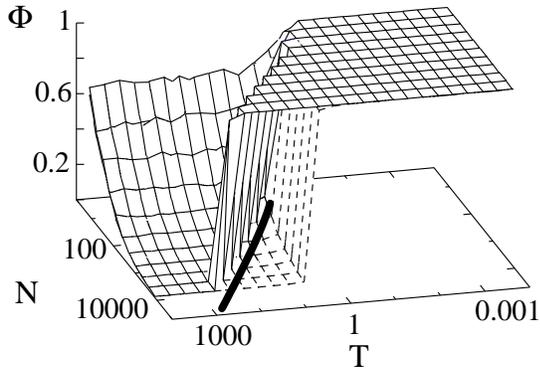}}
\caption{
The order parameter $\Phi=\Phi_k=k_{\rm max}/M$ as a function of
the temperature and the system size for $E=\sum_i-k_i^2/2$ and
$\langle k\rangle=0.5$.
The simulations started either from a star 
(corresponding to $T=0$, solid line)
or a classical random graph 
($T=\infty$, dashed line). 
Each data point represents a single run, time averaged between
$t=100N$ and $200N$ Monte-Carlo steps. 
The thick solid line shows the analytically calculated
spinodal $T_1=M/\ln(N)$.}
\label{MC_k2result}
\end{figure}

For $f(k_i)=-k_i^2/2$, the topology with the lowest overall energy is
a ``star'', where all the $M$ edges are connected to a single node.
Both Monte-Carlo (MC) simulations (see Fig.\ \ref{MC_k2result})
and a simple theoretical
approximation indicate that a first order phase transition occurs
between a dispersed configuration and a star as the temperature is
varied. 
For large enough systems, a sudden change of the order
parameter between zero and one can be observed. The hysteresis appearing
between cooling and heating is consistent with a first order
transition.

The numerical results are supported by a {\it simple theoretical analysis} of
the free energy. The idea is to focus on the regime, where the
most highly connected vertex has already accumulated the majority of
the edges ($\Phi_k>1/2$) and, thus, the energy of the entire graph can be
approximated by $f(k_{\rm max})=-k_{\rm max}^2/2$.
Then, in the thermodynamic limit, the free energy to leading order can
be written as
\be
\f{F(\Phi_k,T)}{MT} \approx
 \f{f(\Phi_k M)}{MT}
 +\Phi_k\ln(N),
\label{F_star}
\ee
where the $\Phi_k$ independent terms have been dropped.
Note that this approximation would be valid even for $\Phi_k<1/2$, if the
energy of the graph was simply defined as $E=f(k_{\rm max})$. 

For $f(\Phi_k M)=-(\Phi_k M)^2/2$ the parabola given by Eq.\
(\ref{F_star}) has a maximum at $\Phi_k=T/M\ln(N)$. When $T\to 0$, this
maximum also shifts towards zero and $F(\Phi_k,T)$ becomes a descending
parabola on the $[0,1]$ interval. This means that the minimum of the
free energy is at $\Phi_k=1$, the star configuration. In contrast, when
the temperature goes above the $T_1=M/\ln(N)$ spinodal point
(thick solid line in Fig.\ \ref{MC_k2result}),
the maximum leaves the $[0,1]$
interval and the free energy becomes an ascending parabola, resulting
in a minimum at a low value of $\Phi_k$ (corresponding to an ER random
graph). However, this value cannot be deduced from Eq.\
(\ref{F_star}), because it is a valid approximation only for
$\Phi_k>1/2$.
For intermediate temperatures the maximum of the parabola separates the
two extreme topologies (the dispersed random graph and the star), among
which one is a metastable configuration and the other one is absolutely
stable.

\begin{figure}[t!] 
\centerline{\includegraphics[angle=-90,width=0.9\columnwidth]{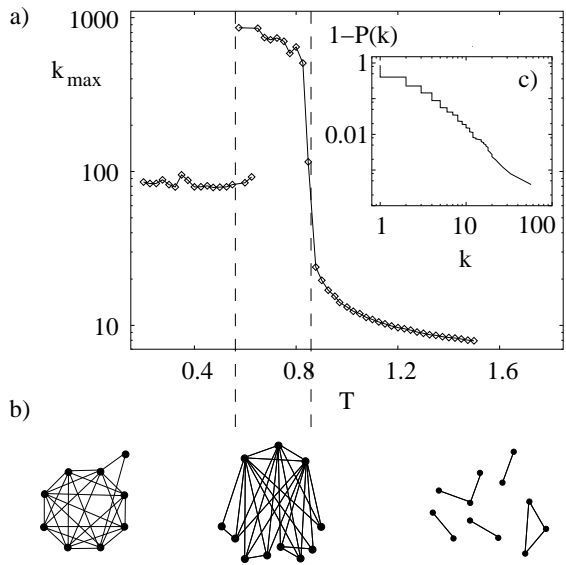}}
\caption{
Phases of the graph when the energy is $E=-\sum_i k_i\ln(k_i)$.
(a) The largest degree $k_{\rm max}$
for $N=10,224$ vertices and $M=2,556$ edges.
Each data point represents a single run, time averaged between
$t=5,000N$ and $20,000N$ MC steps.
The data points are connected to guide the eye.
There is a sharp, continuous transition near
$T=0.85$ and a first-order transition (with a hysteresis) 
around $T=0.5-0.6$.
(b) The three different plateaus in (a)
correspond to distinct topological phases:
$k_{\rm max}={\cal O}(1)$ to the classical random graph,
$k_{\rm max}={\cal O}(M)$ to the star phase
(a small number of stars sharing most of their neighbors)
and 
$k_{\rm max}={\cal O}(\sqrt{M})$ to the fully connected subgraph.
(c) The (cumulative) degree distribution at $T=0.84$ and $t=600N$
follows a power law.
}
\label{MC_kmax}  
\end{figure}

Another application-motivated  choice for the single vertex energy is
$f(k_i)=-k_i\ln(k_i)$, or equivalently, $g(k_i)=-\ln(k_i)$,
inspired, in part, by the logarithmic law of sensation. It is the
logarithm of the degree of a vertex that its neighbors can sense and
benefit from. For this energy the configuration of lowest energy is a
fully connected subgraph [or almost fully connected if $M$ cannot be
expressed as $n(n-1)/2$]. However, the star configuration is also quite
favorable. The order parameter $\Phi_k=k_{\rm max}/M$ can easily
distinguish between these two configurations, because
$k_{\rm max}\approx \sqrt{2M}$ for a fully connected subgraph and
$k_{\rm max}\approx M$ for a star.
Our MC simulations demonstrate (Fig.\ \ref{MC_kmax}) that as we cool
down the system, the dispersed random graph first assembles to a
configuration with a few large stars (sharing most of their neighbors),
and then at lower temperatures it reorganizes into an almost fully
connected subgraph. The hysteresis near this latter transition suggests
that it is a first order phase transition. On the other hand, the former
transition is accompanied by a singularity in the heat capacity and no
hysteresis is observed, indicating that it is a second order phase
transition.

For $\Phi_k>1/2$ Eq.\ (\ref{F_star}) can be used again as a good
approximation for the free energy of the graph. By plugging
$f(\Phi_k M)=-(\Phi_k M)\ln(\Phi_k M)$
into that expression, we get
$F(\Phi_k,T)/(MT) \approx (1-1/T)\ln(N)\Phi_k$
to leading order, which is linear in $\Phi_k$. In agreement with our
observations above, this formula predicts that for $T<1$ the star is a
stable configuration ($\Phi_k$=1 is a minimum of the free energy), and
for $T>1$ it becomes unstable. The transition at $T=T_{\rm c}=1$ is
thus step-like with no hysteresis, indicating a second order phase
transition with an infinitely large critical exponent. The observed
deviation of $T_{\rm c}$ from 1 is a finite size effect.

A remarkable feature of the MC dynamics is that by crossing
$T_{\rm c}$ from above, a scale-free graph (with a degree distribution $\sim k^{-\gamma}$ with $\gamma\simeq 3$) appears at some point of the
evolution of the graph form the random configuration towards the star.
This supports the notion that scale-free graphs are typically
non-equilibrium (dynamical) configurations.
The MC dynamics is governed by the change of the energy associated with
the relocation of an edge. Estimating the energy change of a vertex by
the derivative of the single vertex energy $f(k_i)=-k_i\ln(k_i)$, we
get $\Delta E=1-\ln(k_i)$. Plugging this into the Boltzmann factor,
$\exp[-\Delta E/T]$, at $T=T_{\rm c}=1$ we get a
quantity proportional to $k_i$ for the acception/rejection ratio of a
randomly selected move. Since the preferential attachment in the
Barab\'asi-Albert model
\cite{b-a-science}
is proportional to $k_i$, it is {\it natural that our dynamics
also produces scale-free graphs}.

Due to the macroscopic number of edges of the most highly connected
vertices in the compact configurations, the graph energy is a
non-extensive quantity. More precisely, the energy in different
topological states scales differently with the size of the system.
For instance, when $f(k_i)=-k_i\ln(k_i)$, the energy of the star and the fully
connected subgraph scales as $N\ln(N)$, whereas that of the dispersed
state scales as $N$. Thus (unlike in the mean-field Ising model), there
is {\it no way to choose an appropriate coupling constant that could render
the energy extensive in all topological states simultaneously}.

However, the dispersed state (having an extensive graph energy) can
equally be studied in the grand canonical ensemble. There, the degree
distribution can be expressed as \cite{berg}
$P_k=C\exp[-f(k)/T-\mu k]/k!$,
where C is a normalization factor and the chemical potential $\mu$ is
adjusted to give the correct $\kav$. For $f(k)=-k\ln(k)$, using
Stirling's formula, the distribution takes the form
$P_k=C\exp[-(\mu-1)k]k^{(1/T-1)k}/\sqrt{2\pi k}$.
When $T>1$, this has a tail, which decays faster than exponential,
consequently, each vertex has a small degree. For $T<1$, on the other
hand, the tail becomes divergent, signaling a phase transition at
$T=T_{\rm c}=1$. Note however, that in the $T<1$ temperature range,
due to the non-extensive contribution of the diverging degrees, the
ensembles are not equivalent, and the grand canonical description
looses its validity.

At the critical temperature, the grand canonical description might
still be valid. There, by choosing a more general single vertex energy
$f(k_i)=-(k_i-\alpha)\ln(k_i)$ and setting $\kav$ such that $\mu=1$,
the degree distribution acquires a power law tail
($P_k\sim k^{-(\alpha+1/2)}$) and the network becomes scale-free.
We have to stress though that the scale-free network at $T_{\rm c}$ is
not general: for $\mu>1$ the tail decays exponentially, and for $\mu<1$
the tail diverges.

Although in this paper we assumed that $\kav\leq1$, this is not a
necessary requirement, when the energy is assigned to single vertices.
For large average degree ($\kav>2$) the only difference is that one vertex
cannot collect all the edges, and thus, several stars appear in the
``star'' configuration. Finally, further interesting directions in the
context of the above study include the investigation of additional
relevant forms for the energy [e.g., $E=(k-n)^2$ with $n>1.5$] and the
joint effects of restructuring and growth.

\begin{acknowledgments}
The authors are grateful to G\'abor Tusn\'ady for many valuable
discussions.
This research has been supported by Hungarian National
Science Fundation, grant No: OTKA 034995.
\end{acknowledgments}

\vspace{-5pt}                           %%% DELETE %%%

\end{document}